\pdfoutput=1
\documentclass[preprint,preprintnumbers,amsmath,amssymb,superscriptaddress]{revtex4}
 
\newcommand{\be}{\begin{equation}}
\newcommand{\ee}{\end{equation}}
\newcommand{\ba}{\begin{eqnarray}}
\newcommand{\ea}{\end{eqnarray}}
\newcommand{\bd}{\begin{displaymath}}
\newcommand{\ed}{\end{displaymath}}

\newcommand{\commentout}[1]{{}}

\usepackage{amssymb}
\usepackage{color}
\usepackage{slashed}
\usepackage{amsmath}
\usepackage{graphicx}
\usepackage{epstopdf}
\usepackage{dcolumn}
\usepackage{bm}
\usepackage{hyperref}

\begin{document}

\title{Numerical integration of thermal noise in relativistic hydrodynamics}
\author{Clint Young}
\email{young@physics.umn.edu}
\affiliation{University of Minnesota}
\date{\today}

\begin{abstract}

Thermal fluctuations affect the dynamics of systems near critical points, the evolution of the early universe, and two-particle correlations in heavy-ion collisions. For the latter, numerical 
simulations of nearly-ideal, relativistic fluids are necessary. The correlation functions of noise in relativistic fluids are calculated, stochastic integration of the noise in 3+1-dimensional 
viscous hydrodynamics is implemented, and the effect of noise on observables in heavy-ion collisions is discussed. Thermal fluctuations will cause significant variance in the event-by-event distributions of integrated $v_2$ while changing average values even when using the same initial conditions, suggesting that including thermal noise will lead to refitting of the hydrodynamical parameters with implications for understanding the physics of hot QCD.

\end{abstract}

\maketitle

\section{Introduction}

Any dissipative system in thermal equilibrium must have thermal fluctuations if the system's degrees of freedom have thermal expectation values. This fact is expressed in generality by the 
fluctuation-dissipation relation for bosonic degrees of freedom in thermal equilibrium, 
\begin{equation}
G_S(\omega) = -(1+2n_B(\omega)){\rm Im}(G_R(\omega)){\rm ,} \nonumber
\end{equation}
where $G_S(\omega)$, the Fourier transform of the autocorrelation function, is used to calculate variances of $\hat{\phi}$, and $G_R(\omega)$ is the Fourier transform of the retarded Green 
function, used to calculate the evolution of $\left \langle \delta \hat{\phi}(t) \right \rangle$ in time. 

This relationship determines the fluctuations of macroscopic quantities $e$ and ${\bf v}$ in a viscous fluid \cite{Landau:1980st}. Recently, these fluctuations were determined for relativistic 
fluids in both the Landau-Lifshitz and the Eckart frames \cite{Kapusta:2011gt}, finding the fluctuations of the energy-momentum tensor to have the autocorrelation 
function related to the shear and bulk viscosities by
\begin{equation}
\left \langle \Xi^{ij}(x) \Xi^{kl}(x^{\prime}) \right \rangle = 2\eta T \left[\delta^{ik}\delta^{jl}+\delta^{il}\delta^{jk}\right]+2T\left(\zeta-{\textstyle\frac{2}{3}}\eta \right)\delta^{ij}\delta^{kl} \nonumber
\end{equation}
 in the Landau-Lifshitz frame at rest. This result was applied to 1+1-dimensional boost-invariant hydrodynamics, the approximate description of heavy-ion collisions where 
$\sqrt{s}/A > 100\;{\rm GeV}$. The two-particle correlation $K(\Delta y)$ as a function of the pseudorapidity gap between particles was calculated in boost-invariant hydrodynamics using parameters appropriate for the Large Hadron Collider (LHC). A small but non-zero correlation was found up to $\Delta y=4$, suggesting that measurements of two-particle correlations contain signals of thermal fluctuations from early times of the hydrodynamical evolution of the system.

The authors of \cite{Kapusta:2011gt} point out that thermal fluctuations can be used to determine the viscosity of the fluid produced in heavy-ion collisions. Using fluctuations to determine a 
system's dissipation is not common; more often, the dissipation of a system (for example, the drag of a heavy particle, or the resistivity of a conductor) is the quantity measured directly and is used to determine the corresponding fluctuating quantity at a given temperature. However, heavy-ion collisions are unique dissipative systems of current interest: they exist for $\sim 10\;{\rm fm/c}$ and 
the effect of dissipation is measured relatively indirectly (through collective flow measurements of the produced charged particles) compared with other physical systems. Recently, the role of 
collective flow in determining viscosity has been called into question by measurements of $d+Au$-collisions at the Relativistic Heavy-Ion Collider (RHIC) \cite{Adare:2013piz}. 
The complementary measurement of viscosity through thermal fluctuations might help answer questions concerning thermalization and flow in these experiments.

While these fluctuations might be useful phenomenologically, they present challenges both to the theoretical understanding of hydrodynamics and to numerical simulations. In 
\cite{Kovtun:2011np}, thermal fluctuations lead to an effective viscosity $\eta_{{\rm eff}}$, describing the propagation of sound, as a function of the ``classical" viscosity $\eta_{{\rm cl}}$. A term 
in $\eta_{{\rm eff}}$ is proportional to $1/\eta_{{\rm cl}}^2$, causing a lower bound in the effective viscosity. 
The fluctuations cause divergences which ultimately lead to a breakdown in hydrodynamics. Section \ref{stochastic} presents the numerical implications of these divergences: the variance of the averaged stochastic noises over cells diverges as $1/\sqrt{\Delta V \Delta \tau}$, making exceedingly fine resolutions of thermally fluctuating hydrodynamics nonsensical in its most straightforward implementation, as opposed to dissipative hydrodynamics, where the continuum limit converges. The hydrodynamical limit is saved only upon noticing that many observables, such as the total yields of particles produced in a heavy-ion collision and their elliptic flow, are described well with simulations with coarse grids and are not sensitive to increases in resolution that will break the assumptions of hydrodynamics.

While thermal fluctuations are important, they co-exist in each event with the fluctuations of initial conditions and, depending on the experimental analysis, jet-bulk interactions. Numerical simulation is a practical approach for comparing new theoretical results with experiment without neglecting any of the physics affecting these observables. This paper describes the implementation of thermal noise in a 3+1-dimensional viscous hydrodynamical algorithm. In Section \ref{thermal}, the autocorrelation for noise is determined for Israel-Stewart hydrodynamics. Here, the derivation of the autocorrelation function uses the definitions commonly used in finite-temperature field theory, as was done previously for examining fluctuations in AdS in 
\cite{Son:2009vu}. Some of the same issues examined in Section \ref{thermal} were examined in \cite{Murase:2013tma}; however, Section \ref{thermal} emphasizes that causal stochastic hydrodynamics can be simulated exactly, with {\it white noise} in the the appropriate place in the Israel-Stewart equations, making thermal fluctuations easily simulated with modifications of existing viscous hydrodynamical codes. Section \ref{stochastic} shows how this stochastic process with multiplicative noise can be integrated using a 3+1-dimensional viscous hydrodynamical code for heavy-ion collisions \cite{Dusling:2007gi, Schenke:2010nt, Schenke:2010rr}. Finally, Section \ref{observables} demonstrates which observables will be affected by the presence of thermal noise.

\section{Thermal Green functions and relativistic noise}
\label{thermal}

\subsection{Introductory example: the Langevin equation}
\label{Langevin}

Examining Brownian motion in one dimension illustrates the steps necessary for examining thermal noise in fluids.
The Langevin equation includes a drag force $-\eta p$ linearly proportional to the momentum and a noise term $\xi(t)$ independent of momentum:
\begin{equation}
\frac{dp}{dt} = -\eta p +\xi(t){\rm .}
\end{equation}
The Green function
\begin{equation}
G_R(t-t^{\prime}) = \theta(t-t^{\prime})\exp(-\eta (t-t^{\prime}))
\end{equation}
where $p(t) = \int_0^t dt^{\prime} \; G_R(t-t^{\prime}) \xi(t^{\prime})$. The Fourier transform of $G_R(t)$ can be used to find the autocorrelation function
\begin{eqnarray}
G_S(\omega) &=& -\frac{2T}{\omega}{\rm Im}\{ G_R(\omega)\} \nonumber \\
&= & -\frac{2T}{\omega} \frac{-\omega}{\omega^2+\eta^2} \nonumber \\
&=& \frac{iT}{\eta}\left[ \frac{1}{\omega+i\eta} - \frac{1}{\omega-i\eta} \right] {\rm ,} \\ 
\nonumber
\end{eqnarray}
whose inverse Fourier transform gives $\left \langle p(t) p(t^{\prime}) \right \rangle$ up to normalization:
\begin{equation}
\left \langle p(t) p(t^{\prime}) \right \rangle = A \exp(-\eta |t-t^{\prime}| ){\rm .}
\end{equation}
Remembering $\left \langle p^2(t) \right \rangle = 2MT$, its thermal expectation value, determines $A=2MT$. Finally, 
\begin{equation}
\left \langle \xi(t) \xi(t^{\prime}) \right \rangle = \left \langle (\dot{p}(t)+\eta p(t)) (\dot{p}(t^{\prime})+\eta p(t^{\prime})) \right \rangle = 2\eta MT \delta(t-t^{\prime}){\rm ,}
\end{equation}
when $p$ and $\xi$ are uncorrelated, determine the Einstein relation between drag and noise. In summary, the response of the heavy particle to the stochastic force, and the fact that 
the heavy particle exists in thermal equilibrium, is enough to determine the autocorrelation function of $\xi(t)$.

\vspace{.25cm}

\subsection{Thermal noise in hydrodynamics}

We work in the Landau-Lifshitz frame, 
where $u^{\mu}$ is defined as the flow of energy density in the fluid, normalized to 1. The Israel-Stewart form of causal hydrodynamics, driven by a noise term $\Xi^{\mu \nu}(x)$, is 
\begin{eqnarray}
\partial_{\mu} T^{\mu \nu}_{ {\rm id.} } & = & -\partial_{\mu} W^{\mu \nu} - \partial_{\mu} \Xi^{\mu \nu}{\rm ,} \nonumber \\
\left( u\cdot \partial \right) W^{\mu \nu} & = & -\frac{1}{\tau_{\pi}} \left( W^{\mu \nu}-\eta \Delta^{ \langle \mu} u^{\nu \rangle} - \zeta (\partial \cdot u)\Delta^{\mu \nu} \right) {\rm ,} \\
\nonumber
\label{ISNoise1}
\end{eqnarray}
where $T^{\mu \nu}_{ {\rm id.} }= -pg^{\mu \nu}+(e+p)u^{\mu}u^{\nu}$, $\Delta^{\mu} \equiv \partial^{\mu}-u^{\mu}(u\cdot \partial)$, $\Delta^{\mu \nu} \equiv u^{\mu} u^{\nu}-g^{\mu \nu}$, 
and $\Delta^{ \langle \mu} u^{ \nu \rangle} = \Delta^{\mu}u^{\nu}+\Delta^{\nu}u^{\mu}-\frac{2}{3}g^{\mu \nu} (\partial \cdot u)$. At this point, the equations are not closed, because the noise 
$\Xi^{\mu \nu}$ has not yet been specified. However, the fluctuation-dissipation relation determines the noise, as will now be demonstrated.

Without any loss of generality, we examine the case of a fluid at rest. The equations can be linearized by substituting $u=(1, {\bf 0})+\delta u$, $e=e_0+\delta e$, and $p=p_0+\left( \frac{\partial p}{\partial e} \right) \delta e$. The homogeneous equations are 
\begin{equation}
\partial_t \delta e = (e_0+p_0){\bf \nabla}\cdot {\bf \delta u}{\rm ,} \nonumber
\end{equation}

\vspace{-.5cm}

\begin{equation}
(e_0+p_0) \partial_t \delta u^i - \partial_i \left( \frac{\partial p}{\partial e} \right) \delta p + \partial_j W^{ij} = 0 {\rm ,} \nonumber
\end{equation}

\vspace{-.5cm}

\begin{equation}
\partial_t W^{ij} = -\frac{1}{\tau_{\pi} } \left( W^{ij} + \eta \partial^{ \langle i} \delta u^{j \rangle} +\zeta(\nabla \cdot {\bf u})\delta^{ij} \right) {\rm .}
\end{equation}
The limit of $\tau_\pi \to 0$ is acausal but instructive. In this limit, $W^{\mu \nu} \to  -\eta \partial^{ \langle i} \delta u^{j \rangle} -\zeta(\nabla \cdot {\bf u})\delta^{ij}$.
We work in Fourier space and separate the velocity perturbations into parts $\delta u = \delta u_L + \delta u_T$, where 
$\delta u_L^i(\omega, {\bf k}) \equiv {\bf k} \cdot {\bf \delta u}/|{\bf k}|$ and $\delta u_T(\omega, {\bf k}) \equiv \delta u(\omega, {\bf k}) - \delta u_L(\omega, {\bf k})$. For $\delta u_T$, this reduces to 
a single equation:
\begin{equation}
\left[ -i(e_0+p_0) \omega + \eta |{\bf k}|^2 \right]\delta u_T^i = 0{\rm .}
\end{equation}
The longitudinal component of velocity is coupled to the pressure, but a single equation of motion can be determined with some algebraic substitutions:
\begin{equation}
\left[ -i(e_0+p_0) \omega +i \frac{ (\frac{\partial p}{\partial e} ) }{\omega} |{\bf k}|^2 + (\zeta +{\textstyle\frac{4\eta}{3}}) |{\bf k}|^2 \right] \delta u_L^i = 0{\rm .}
\end{equation}
The retarded Green function of $\delta u^i$ for perturbations to the energy-momentum tensor $\delta T^{0 j}$ is 
\begin{equation}
G^{ij}_R(\omega, {\bf k}) = \frac{\omega}{w}\left[ \frac{ 1 }{\omega + i\frac{\eta}{w}|{\bf k}|^2} \right]\left(\delta^{ij}-\frac{k^i k^j}{|{\bf k}|^2} \right)
+ \frac{\omega}{w} \left[ \frac{1}{\omega - \frac{( \frac{\partial p}{\partial e} )}{\omega} |{\bf k}|^2 + i\frac{\zeta+4\eta /3}{w}|{\bf k}|^2} \right]\frac{k^i k^j}{|{\bf k}|^2}{\rm ,}
\end{equation}
where $w=e_0+p_0$ is the unperturbed enthalpy. If $\Xi^{\mu \nu}(x)$ leads to thermal expectation values for the perturbations, then the autocorrelation function in the rest frame 
\begin{equation}
A^{ik}(x, x^{\prime}) \equiv \left \langle \partial_{j} \Xi^{ij}(x) \partial_{l} \Xi^{kl}(x^{\prime}) \right \rangle = \left \langle \partial_{\mu} (-T^{i \mu}_{ {\rm id.} }(x) - W^{i \mu}(x)) \partial_{\rho}
(-T^{k \rho}_{ {\rm id.} }(x^{\prime}) - W^{k \rho}(x^{\prime})) \right \rangle 
\end{equation}
can be determined using the fluctuation-dissipation relation. In Fourier space,
\begin{eqnarray}
A^{ik}(\omega, {\bf k}) &=& -\frac{2T}{\omega}(-iw\omega+\eta |{\bf k}|^2)(+iw\omega+\eta|{\bf k}|^2){\rm Im} \left\{ \frac{\omega/w}{\omega+i \frac{\eta}{w}|{\bf k}|^2} \right\} 
\left[ \delta^{ik}-\frac{k^i k^k}{|{\bf k}|^2} \right] \nonumber \\
& &  -\frac{2T}{\omega}\left((w\omega-\frac{(\frac{\partial p}{\partial e})}{\omega}|{\bf k}|^2)^2+((\zeta+\frac{4}{3}\eta)|{\bf k}|^2)^2 \right) \nonumber \\
& & \times {\rm Im} \left\{ \frac{\omega/w}{\omega - \frac{( \frac{\partial p}{\partial e} )}{\omega} |{\bf k}|^2 + i\frac{\zeta+4\eta /3}{w}|{\bf k}|^2} \right\} \frac{k^i k^k}{|{\bf k}|^2} \nonumber \\
&=& 2\eta T \left[ \delta^{ik} |{\bf k}|^2-k^i k^k \right] + 2({\textstyle\frac{4 \eta}{3}}+\zeta)T k^i k^k {\rm ,} \\
\nonumber
\end{eqnarray}
making
\begin{equation}
A^{ik}(x, x^{\prime}) = \left[ 2\eta T [\delta^{ik}\nabla^2-\partial^i \partial^k] + 2(\zeta+{\textstyle\frac{4\eta}{3}})T\partial^i \partial^k \right] \delta^4(x-x^{\prime}) {\rm .}
\label{Aik}
\end{equation}
Now, consider $G^{ijkl} \equiv \left \langle \Xi^{ij}(x) \Xi^{kl}(x^{\prime}) \right \rangle$. The symmetry condition $\Xi^{ij} = \Xi^{ji}$ and Equation \ref{Aik} are enough to determine 
\begin{equation}
G^{ijkl} = \left[ 2\eta T(\delta^{ik}\delta^{jl}+\delta^{il}\delta^{jk}) + 2(\zeta-2\eta/3)T \delta^{ij}\delta^{kl} \right] \delta^4(x-x^{\prime}){\rm .}
\end{equation}

To determine the autocorrelation of noise when $\tau_{\pi} \ne 0$, note that 
\begin{equation}
W^{ij}(\omega, {\bf k}) = \frac{ -\eta \partial^{ \langle i} \delta u^{j \rangle} - \zeta ({\bf \nabla} \cdot \delta {\bf u}) \delta^{ij}}{1-i\tau_{\pi}\omega} {\rm .}
\end{equation}
This simply changes the homogeneous equations of motion above to 
\begin{equation}
\left[ -i(e_0+p_0) \omega + \frac{\eta |{\bf k}|^2}{1-i\tau_\pi \omega} \right]\delta u_T^i = 0{\rm ,} \nonumber 
\end{equation}

\vspace{-0.5cm}

\begin{equation}
\left[ -i(e_0+p_0) \omega +i \frac{ (\frac{\partial p}{\partial e} ) }{\omega} |{\bf k}|^2 + (\zeta +{\textstyle\frac{4\eta}{3}}) \frac{|{\bf k}|^2}{1-i\tau_\pi \omega} \right] \delta u_L^i = 0{\rm .}
\end{equation}
Skipping some of the same steps as above, this leads to a new autocorrelation function $A^{\prime ik}(\omega, {\bf k})$, given by 
\begin{equation}
A^{\prime ik}(\omega, {\bf k}) = \frac{A^{ik}(\omega, {\bf k})}{1+(\tau_\pi \omega)^2} {\rm ,}
\end{equation}
so that now
\begin{equation}
\left \langle ( (\tau_\pi \partial_t \Xi^{ij}(x) +\Xi^{ij}(x) )(\tau_\pi \partial_t \Xi^{kl}(x^{\prime}) +\Xi^{kl}(x^{\prime}) ) \right \rangle \nonumber
\end{equation}

\vspace{-.5cm}

\begin{equation}
= \big[ 2\eta T(\delta^{ik}\delta^{jl}+\delta^{il}\delta^{jk}) + 2(\zeta-2\eta/3)T \delta^{ij}\delta^{kl} \big] \delta^4(x-x^{\prime}){\rm .}
\label{Aprime}
\end{equation}
This defines a stochastic equation of motion for $\Xi^{\mu \nu}(x)$, 
\begin{equation}
\partial_t \Xi^{ij} = -\frac{1}{\tau_\pi }(\Xi^{ij}-\xi^{ij}){\rm ,}
\label{XiSecondOrder}
\end{equation}
where $\xi^{ij}$ has the same autocorrelation function as the right-hand side of Eq. \ref{Aprime}. Comparing with Section \ref{Langevin}, 
the thermal fluctuation $\Xi^{\mu \nu}$ is analogous to the momentum $p$ in the 
Langevin equation, and $\xi^{\mu \nu}$ is analogous to the noise in the Langevin equation. Finally, we define $W^{\prime}(x) \equiv W^{\mu \nu}(x)+\Xi^{\mu \nu}(x)$. The analysis in 
the fluid at rest is finished; the shift back to relativistic hydrodynamical equations in the Landau-Lifshitz frame is simple, using $\partial_t \to u \cdot \partial$, $\nabla \cdot \delta {\bf u} \to 
- \partial \cdot u$, $\delta^{ij} \to \Delta^{\mu \nu}$, and $\partial^i \delta u^j \to -\Delta^{\mu} u^{\nu}$:

\begin{equation}
\partial_{\mu} T^{\mu \nu}_{ {\rm id.} } = -\partial_{\mu} W^{\prime \mu \nu}{\rm ,}
\nonumber
\end{equation}

\vspace{-.5cm}

\begin{equation}
(u \cdot \partial) W^{\prime \mu \nu} = -\frac{1}{\tau_\pi}\left( W^{\prime \mu \nu}-\eta \Delta^{\langle \mu} u^{\nu \rangle}-\zeta (\partial \cdot u)\Delta^{\mu \nu} -\xi^{\mu \nu} \right){\rm ,}
\nonumber
\end{equation}

\vspace{-.5cm}

\begin{equation}
\left \langle \xi^{\mu \nu}(x) \xi^{\rho \sigma}(x^{\prime}) \right \rangle = \left[2\eta T(\Delta^{\mu \rho}\Delta^{\nu \sigma}+\Delta^{\mu \sigma} \Delta^{\nu \rho})
+2(\zeta-2\eta/3)T\Delta^{\mu \nu} \Delta^{\rho \sigma}\right] \delta^4(x-x^{\prime}){\rm .}
\label{ISnoise}
\end{equation}
Eqs. \ref{ISnoise}, with the equation of state $p(e)$, are now closed; usage of the fluctuation-dissipation relation has determined the autocorrelation function for the thermal noise. 

\section{Stochastic integration of hydrodynamical noise}
\label{stochastic}

Eqs. \ref{ISnoise} describes a {\it stochastic process}: $T_{\rm id.}$ and $W^{\prime}$ are now random variables, and a large ensemble of these tensors will approximate a distribution 
functional. They are described with a partial differential equation involving another stochastic process, $\xi$, called the {\it noise}. The noise is {\it multiplicative}: it depends on $\eta$ and 
$T$, which are functions of $T_{ \rm id}$. \footnote{ For more on stochastic processes, see Adam Monahan's lectures at
http://www.pims.math.ca/scientific/summer-school/summer-school-stochastic-and-probabilistic-methods-atmosphere-ocean-and-cli }.

For familiar, well-behaved functions, the Riemann-Stieljes integral 
\begin{equation}
\int_a^b f\; dg \equiv \lim_{n \to \infty} \sum_{i=1}^n f(\tau_i)(g(t_i)-g(t_{i-1})){\rm ,}
\end{equation}
where the interval $[a,b]$ is partitioned into intervals $\{ [t_{i-1}, t_i]\}$, is well-defined and has the same value for any choices of $t_{i-1}<\tau_i<t_i$. A stochastic process is less well-behaved, 
and as a result, different choices for $\tau_i$ lead to different values of the integral.

To see this, consider $W(t) = \int_0^t dt^{\prime} \xi(t^{\prime})$. This is a Wiener process: a continuous generalization of a random walk where $\left \langle W(t) \right \rangle = 0$ and 
$\left \langle (W(t)-W(t^{\prime}))^2 \right \rangle = t-t^{\prime}$ for any $t$ and $t^{\prime}$. If you choose $\tau_i=t_{i-1}$ (the Ito integral) and define $W_i \equiv W(t_i)$, the integral 
$(Ito)\int_0^t dt^{\prime}\; W(t^{\prime})dW$ becomes 
\begin{eqnarray}
 \lim_{n \to \infty} \sum_{i=1}^n W_{i-1}(W_i-W_{i-1}) = \lim_{n \to \infty} \sum_{i=1}^n  \frac{1}{2}\left[(W_i+W_{i-1})-(W_i-W_{i-1}) \right](W_i-W_{i-1})  \nonumber \\
 =\lim_{n \to \infty} \sum_{i=1}^n \frac{1}{2} \left[ (W_i)^2-(W_{i-1})^2 - \langle (W_i-W_{i-1})^2 \rangle \right] = \frac{1}{2} \left[ W^2(t)-W^2(0)-t \right]{\rm .} \\ \nonumber
 \end{eqnarray}
Choosing $\tau_i=t_{i-1/2} \equiv (t_{i-1}+t_i)/2$ (the Stratonovich integral) leads to an alternating series of cancellations of expectation values so that 
\begin{eqnarray}
 (S)\int_0^tW(t^{\prime})dt^{\prime} &=& \lim_{n \to \infty} \sum_{i=1}^n W( t_{i-1/2} )\left(W(t_i)-W(t_{i-1})\right)  \nonumber \\
 & = & \lim_{n \to \infty} \sum_{i=1}^n \Big(W( t_{i-1/2} )\big(W(t_i)-W(t_{i-1/2})\big)-W( t_{i-1/2} )\big(W(t_{i-1/2})-W(t_{i-1})\big)\Big) \nonumber \\
& = & \lim_{n \to \infty} \sum_{i=1}^n \frac{1}{2}\Big[W^2(t_i)-W^2(t_{i-1/2})-(W(t_i)-W(t_{i-1/2}))^2  \nonumber \\
 & & +W^2(t_{i-1/2})-W^2(t_{i-1})+(W(t_{i-1/2})-W(t_{i-1}))^2 \Big] \nonumber \\
& = & \frac{1}{2}\Big[W^2(t)-W^2(0)\Big]{\rm ,} \nonumber \\
\end{eqnarray}
which differs from the Ito integral for all non-zero values of $t$. 

This ambiguity in defining the integral is analogous to the problem of regulating composite local operators in quantum field theory; there, different expectation values are possible, depending on 
how limits are taken for points in spacetime to be equal. The resolution of the ambiguity in quantum field theory comes from requiring classical equations of motion for the expectation values to 
be true quantum-mechanically as well as classically. The resolution here is related: because $W(t)$ is the approximation of a continuous and differentiable function, the usual rules for integration should be satisfied. This makes the Ito integral unacceptable, with its extra term $-t/2$ in $(Ito)\int_0^t dt^{\prime}\; W(t^{\prime})dW$. The Stratonovich integral has been shown to reproduce the rules of integration obeyed by differentiable functions, in the case of multiplicative noise linearly proportional to $W(t)$. The same can be shown for more complicated functional dependences on $W$.

While the Stratonovich integral is required for this and many other physical systems, Stratonovich integration is somewhat more difficult to implement numerically than is the Ito integral: for a 
differential equation $dX = a(X,t)dt+b(X,t)\xi(t)dt$, the Ito integral is approximated by Euler's method
\begin{equation}
X^{n+1}=X^n+a(X^n,t^n)\Delta t+b(X^n, t^n)\xi^n \Delta t{\rm ,}
\end{equation}
by definition of the Ito integral. Here, $\xi^n \equiv \frac{1}{\Delta t} \int_{t_i}^{t_{i+1} } \xi(t^{\prime})dt^{\prime}$ is stochastic and in practice is randomly sampled. Oftentimes, when the 
Stratonovich integral of a stochastic process is needed, it is recast as an Ito integral using drift-correction terms, which can be determined by requiring the two different integrals to reach 
the same thermal expectation values \cite{Arnold:1999uza}. For Eqs. \ref{ISnoise}, the drift-correction terms would require determination of partial derivatives of the equation of state, 
prohibitive for numerical integration.

Fortunately, Heun's method, 
\begin{equation}
\bar{X}^{n+1}=X^n+a(X^n,t^n)\Delta t+b(X^n, t^n)\xi^n \Delta t{\rm ,} \nonumber
\end{equation}
\begin{equation}
X^{n+1}=X^n+\frac{1}{2}(a(X^n,t^n)+a(\bar{X}^{n+1},t^{n+1}))\Delta t + \frac{1}{2}(b(X^n,t^n)+b(\bar{X}^{n+1},t^{n+1})) \xi^n \Delta t {\rm ,}
\label{Heun}
\end{equation}
approximates the Stratonovich integral: the averaged values of $a(X,t)$ and of $b(X,t)$ in Eq. \ref{Heun} approximate $a$ and $b$ at the midpoints with a variance averaging to zero.
In fact, Heun's method can outperform other numerical integrations of the Stratonovich integral \cite{GarciaAlvarez:2011}. By using 
Heun's method, the derivatives needed for Euler's method to approximate the Stratonovich integral can be avoided.

\textsc{music} \cite{Schenke:2010nt} can be used to integrate the Israel-Stewart equations : at each timestep, the equations for $T^{0 \mu}_{ \rm id.}$ are solved in $\tau-\eta$ coordinates, using the Kurganov-Tadmor method \cite{Kurganov:2000ne}, which is second-order for smooth flows, switches to first-order when large gradients exist, is conservative, and is well-behaved when 
$\Delta \tau \to 0$. The energy density $e$ and velocity $u^{\mu}$ is reconstructed, and used to determine $T^{ij}_{ \rm id.}$. $W^{\mu \nu}$ also is determined using the Kurganov-Tadmor method. Importantly, \textsc{music} can use either Euler's method or Heun's method.

$\Xi^{\mu \nu}$ is solved separately: Eq. \ref{XiSecondOrder} is written
\begin{equation}
(S)\Delta \Xi^{\mu \nu} = \left[ -{\bf u} \cdot (\nabla \Xi^{\mu \nu}) - \frac{1}{\tau_\pi}(\Xi^{\mu \nu}-\xi^{\mu \nu}) \right] \Delta \tau/u^0 {\rm ,}
\label{XiDiscrete}
\end{equation}
where $(S)$ is a reminder that this differential equation is solved with the Stratonovich integral. The noise term $\xi^{\mu \nu}$ is multiplicative and therefore changes with each Runge-Kutta step as does 
the estimate for $\eta$ and $T$. 

Eq. \ref{XiDiscrete} must be worked out in $\tau-\eta$ coordinates: the metric $g_{\mu \nu} = {\rm diag}(1,-1,-1,-\tau^2)$ leads to non-trivial Christoffel symbols. For any tensor $T^{\mu \nu}$, 
\begin{equation}
T^{\tau \eta}_{; \tau}=T^{\tau \eta}_{, \tau}+\frac{1}{\tau}T^{\tau \eta}{\rm ,}\;\; T^{i \eta}_{; \tau}=T^{i \eta}_{, \tau}+\frac{1}{\tau}T^{i \eta}{\rm ,}\;\;
T^{\eta \eta}_{; \tau}=T^{\eta \eta}_{, \tau}+\frac{2}{\tau}T^{\tau \eta}{\rm .}
\end{equation}
Defining
\begin{equation}
\tilde{T}^{\mu \nu} \equiv 
\begin{pmatrix} T^{\tau \tau} & T^{\tau j} & \tau T^{\tau \eta} \\ 
T^{i \tau} & T^{i j} & \tau T^{i \eta} \\ 
\tau T^{\tau \eta} & \tau T^{\eta j} & \tau^2 T^{\eta \eta} \end{pmatrix}
\end{equation}
leads to the simplified expression
\begin{equation}
\tilde{T}^{\mu \nu}_{; \tau} = \tilde{T}^{\mu \nu}_{, \tau}{\rm .}
\end{equation}
The remaining covariant derivatives are 
\begin{equation}
\tilde{T}^{\mu \nu}_{; i} = \tilde{T}^{\mu \nu}_{, i}{\rm ,}\;\; \tilde{T}^{\mu \nu}_{; \eta} = \tilde{T}^{\mu \nu}_{, \eta}+\left( \delta^{\mu \tau} \tilde{T}^{\eta \nu} + \delta^{\nu \tau} \tilde{T}^{\mu \eta} \right)
+\left( \delta^{\mu \eta} \tilde{T}^{\tau \nu} + \delta^{\nu \eta} \tilde{T}^{\mu \tau} \right){\rm .}
\end{equation}
This is used to determine $-{\bf u}^i T^{\mu \nu}_{; i}=-\tilde{\bf u}^i \tilde{T}^{\mu \nu}_{; i}$, where $\tilde{u}$ is defined $\tilde{u}\equiv (u^{\tau}, u^x, u^y, u^{\eta}/\tau)$.

Finally, the noise must be sampled for each step in $\tau$. In the rest frame of the fluid, define the averaged noise 
\begin{equation}
\Delta \xi^{ij}_{\alpha} \equiv \frac{1}{\Delta V \Delta \tau}\int_{ {\rm cell}} d^4x \; \xi^{ij}(x)
\end{equation}
so that $\alpha$ labels the indices in the discretized space and time of any simulation. For a cell of size $\Delta V \Delta \tau$, 
\begin{equation}
\left \langle \Delta \xi^{ij}_{\alpha} \Delta \xi^{kl}_{\alpha^\prime} \right \rangle = \delta_{\alpha \alpha^\prime} 2\eta_sT \frac{1}{(\Delta V \Delta \tau)^2} 
\left[ \delta^{ik}\delta^{jl}+\delta^{il}\delta^{jk}-2\delta^{ij}\delta^{kl}/3 \right]
\Delta V \Delta \tau {\rm .}
\end{equation}
Here, $\eta_s$ now signifies the shear viscosity to avoid confusion with the rapidity coordinate, and the bulk viscosity $\zeta=0$. 
The symmetry of $\xi^{ij}$ requires the correlation function to have the structure $A(\delta^{ik} \delta^{jl}+\delta^{il}\delta^{jk})+B\delta^{ij}\delta^{kl}$. 
Although $\delta_{\alpha \alpha^\prime}$ forces there to be no correlation of the noise terms between different cells, there is a non-trivial correlation between the diagonal terms of 
$\Delta \xi^{ij}$. This correlation is simplified upon noting that the autocorrelation of traces $\left \langle \Delta \xi^{ii}_{\alpha} \Delta \xi^{jj}_{\alpha^\prime} \right \rangle = 0$; for any tensor 
$\tilde{\xi}^{ij}$ with the autocorrelation $\left \langle \tilde{\xi}^{ij} \tilde{\xi}^{kl} \right \rangle = A[\delta^{ik}\delta^{jl}+\delta^{il}\delta^{jk}]$, 
\begin{equation}
\left \langle (\tilde{\xi}^{ij} - \delta^{ij} \tilde{\xi}^{aa}/3)(\tilde{\xi}^{kl} - \delta^{kl} \tilde{\xi}^{bb}/3) \right \rangle = A\left[ \delta^{ik}\delta^{jl}+\delta^{il}\delta^{jk}-2\delta^{ij}\delta^{kl}/3 \right]{\rm .}
\end{equation}
This suggests a simple procedure for sampling traceless symmetric tensors with the correlation function needed: simply sample a symmetric tensor, and subtract one third of the trace from the 
diagonal elements.

If the coarse-graining is large compared to the correlation lengths of the microscopic theory which is approximated by these equations, then the central limit theorem can be applied to the fluctuations of this system, and the distribution function for $\Delta \xi^{ij}_\alpha$ is approximated by a Gaussian distribution. This large coarse-graining is possible when all gradients $|\nabla e/e|$ are small compared to the inverse of the mean-free path $1/\lambda$, the same condition necessary for any hydrodynamical system to be accurate. 

However, the averaged noise diverges with decreasing cell sizes as $1/\sqrt{\Delta V \Delta \tau}$. While the cells always have finite sizes, this divergence causes problems for 
hydrodynamical equations themselves, because of the large gradients created by decreasing the cell size. The relaxation time $\tau_\pi$ regulates this divergence in time but not in volume. 
Through the transport coefficients of $\eta_s$ and $\tau_\pi$, the microscopic theory itself sets a minimum scale for the accuracy of thermally fluctuating hydrodynamics. 

One may notice at this point that this divergence does not cause problems for the analytic calculations of $K(\Delta y)$ in \cite{Kapusta:2011gt}. Indeed, the same approach of separating perturbations from the hydrodynamics can be implemented numerically: the equations can be separated into an equation for the unperturbed background and an equation for the noise and its response. However, this approach ultimately ignores the real problem of resolution in thermally fluctuating hydrodynamics that has been encountered.

At this point, progress can be made by noticing that the data from RHIC and the LHC is not highly resolved in pseudorapidity or in azimuth. Quantities such as the spectra of charged light 
hadrons and $v_2$ are determined with hydrodynamics without overly high resolutions, thanks to the large system sizes produced in heavy-ion collisions.

This completes the description of numerical integration of thermal noise in heavy-ion collisions. While simulations including thermal noise is necessary, some back-of-the-envelope estimates of 
the effect of thermal noise are now needed.

\section{Effects of thermal noise on observables}
\label{observables}

Here is a good point to take a comprehensive look at the observables and scales of heavy-ion collisions to determine which measurements are affected by thermal noise and by how much. 
The expectation value $\left \langle \Xi^{\mu \nu}(x) \right \rangle = 0$: the expectation value of the energy-momentum tensor $\left \langle T^{\mu \nu}(x) \right \rangle$ is unaffected by thermal 
noise and is the same as the result from a single calculation without noise and the same initial conditions. For this reason, the spectra of produced hadrons averaged over many collisions is 
unaffected by the presence of noise. However, because $\left \langle \Xi^{\mu \nu}(x) \Xi^{\rho \sigma}(x^{\prime}) \right \rangle \ne 0$, 
$\left \langle T^{\mu \nu}(x) T^{\rho \sigma}(x^{\prime}) \right \rangle \ne 0$ and two-particle correlations will have some non-zero contribution from thermal noise, as was shown in 
 \cite{Kapusta:2011gt}. Because the event plane angles $\Psi_n$ are defined in each collision with averages over measured particles, the quantities $v_n$ are themselves integrals over 
 two-particle correlations and are also sensitive to thermal noise. One can realize their effect on average values by considering a collision with impact parameter $b=0$: in the calculation without 
 noise, all coefficients $v_n=0$. However, almost every calculation sampling the noise will have significant non-zero $v_n$, making every $\left \langle v_n \right \rangle > 0$.
 
What variance of $v_n$ in heavy-ion collisions is expected? One can estimate this at the LHC by examining the energy and length scales of a central lead-lead collision at 
$\sqrt{s}/A = 2.76\;{\rm TeV}$: the energy density at the center of the transverse plane is approximately $132\;{\rm GeV/fm^3}$, and using the equation of state in 
\cite{Huovinen:2009yb}, the entropy density 
$s=337\;{\rm fm^{-3}}$, $p=41\;{\rm GeV/fm^3}$, and $T=513\;{\rm GeV}$. For $v_2$ integrated in $p_T$ and in pseudorapidity in the range $|\eta| < 1$, the same resolution in spacetime rapidity 
is needed, and resolutions of about $0.5\;{\rm fm}$ in the directions transverse to the beam are necessary for determining the elliptic flow accurately in this small system. If the thermalization 
of the hydrodynamical system occurs at approximately $0.5\;{\rm fm}$ and the shear viscosity $\eta_s/s = 0.08$, each component of the cell-averaged $\Xi^{ij}$ has a root mean square of 
$\approx \sqrt{2\eta_sT/(\Delta V\Delta \tau)} = 6.6 \;{\rm GeV/fm^3}$. Comparing this to $p$ suggests that the variance of flow and $v_2$ caused by thermal noise in the most central class of 
lead-lead collisions at the LHC may be on the order of 15\%.

\section{Conclusions}

The fluctuations in a thermal fluid are related to the fluid's transport coefficients. Specifically, the viscosity determines the fluctuations of the energy-momentum tensor describing the fluid. 
Numerical simulation of thermal noise in fluids is possible with only minor modifications of existing viscous hydrodynamical algorithms. However, the modified code is now limited with a 
minimum resolution, related to the physical limit of hydrodynamics which has always been known but was previously ignorable in dissipative hydrodynamics without fluctuations. This is 
despite the fact that the correlation function for thermal noise has been calculated in the linearizable limit of hydrodynamics. Including transport coefficients beyond the viscosities and 
relaxation times might regulate the large gradients introduced by noise, but this has not been examined here.

Analytic calculations show that for the hot, yet small, systems produced in heavy-ion collisions, thermal fluctuations have some measurable effects on observables. This paper has demonstrated 
this effect without yet making comparisons with data. In the various observables in heavy-ion collisions, the thermal fluctuations may prove to be relatively quiet compared with other event-by-event fluctuations. However, because of the the thermal fluctuations' close relationship to transport coefficients, understanding and measuring the effect of these fluctuations will provide important 
insights and independent measurements of the surprising fluid produced in heavy-ion collisions.

\section{Acknowledgments}

This work was supported by the U.S. DOE Grant No. DE-FG02-87ER40328. I especially thank Joseph Kapusta, Todd Springer, Gabriel Denicol, Charles Gale, 
Sangyong Jeon, and Bj\"orn Schenke for helpful discussions and comments.


\begin{thebibliography}{9}

\bibitem{Landau:1980st}
   L.~D.~Landau and E.~M.~Lifshitz, {\bf Statistical Physics: Part 2}
   (Pergamon, Oxford, 1980).

\bibitem{Kapusta:2011gt} 
  J.~I.~Kapusta, B.~Muller and M.~Stephanov,
  Phys.\ Rev.\ C {\bf 85}, 054906 (2012)
  [arXiv:1112.6405 [nucl-th]].
  
\bibitem{Adare:2013piz} 
  A.~Adare {\it et al.}  [PHENIX Collaboration],
  [arXiv:1303.1794 [nucl-ex]].

\bibitem{Son:2009vu} 
  D.~T.~Son and D.~Teaney,
  JHEP {\bf 0907}, 021 (2009)
  [arXiv:0901.2338 [hep-th]].

\bibitem{Kovtun:2011np} 
  P.~Kovtun, G.~D.~Moore and P.~Romatschke,
  Phys.\ Rev.\ D {\bf 84}, 025006 (2011)
  [arXiv:1104.1586 [hep-ph]].

\bibitem{Murase:2013tma} 
  K.~Murase and T.~Hirano,
  arXiv:1304.3243 [nucl-th].

\bibitem{Dusling:2007gi} 
  K.~Dusling and D.~Teaney,
  Phys.\ Rev.\ C {\bf 77}, 034905 (2008)
  [arXiv:0710.5932 [nucl-th]].

\bibitem{Schenke:2010nt} 
  B.~Schenke, S.~Jeon and C.~Gale,
  Phys.\ Rev.\ C {\bf 82}, 014903 (2010)
  [arXiv:1004.1408 [hep-ph]].
  
\bibitem{Schenke:2010rr} 
  B.~Schenke, S.~Jeon and C.~Gale,
  Phys.\ Rev.\ Lett.\  {\bf 106}, 042301 (2011)
  [arXiv:1009.3244 [hep-ph]].

\bibitem{Arnold:1999uza} 
  P.~B.~Arnold,
  Phys.\ Rev.\ E {\bf 61}, 6091 (2000)
  [hep-ph/9912208].

\bibitem{GarciaAlvarez:2011}
   D.~Garc\'ia-\'Alvarez,
   [arXiv:1102.4401[comp-ph]].

\bibitem{Kurganov:2000ne}
   A.~Kurganov and E.~Tadmor,
   Journal of Computational Physics {\bf 160}, 241 (2000)

\bibitem{Huovinen:2009yb} 
  P.~Huovinen and P.~Petreczky,
  Nucl.\ Phys.\ A {\bf 837}, 26 (2010)
  [arXiv:0912.2541 [hep-ph]].

\end{thebibliography}
\end{document}